# A NEUROSCIENCE APPROACH REGARDING STUDENT ENGAGEMENT IN THE CLASSES OF MICROCONTROLLERS DURING THE COVID19 PANDEMIC

**Iuliana MARIN[1]**

[1]*University Politehnica of Bucharest, Faculty of Engineering in Foreign Languages (ROMANIA), marin.iulliana25@gmail.com*

## Abstract

The process of teaching has been greatly changed by the COVID-19 pandemic. It is possible that studying will not resemble anymore the process known by the previous generations of students. As the current generations learn by doing and use their intuition, new platforms need to be involved in the teaching process. The current paper proposes a new method to keep the students engaged while learning by involving neuroscience during the classes of Microcontrollers. Arduino and Raspberry Pi boards are studied at the course of Microcontrollers using online simulation environments. The Emotiv Insight headset is used by the professor during the theoretical and practical hours of the Microcontrollers course. The analysis performed on the brainwaves generated by the headset provides numerical values for the mood, focus, stress, relaxation, engagement, excitement and interest levels of the professor.

The approaches used during teaching were inquiry-based learning, game-based learning and personalized learning. In this way, professors can determine how to improve the connection with their students based on the use of technology and virtual simulation platforms. The results of the test show that the game-based learning was be best approach because students had to become problem solves and start to use the software skills which they will need as future software engineers. The emphasis is put on mastering the mindset by having to choose their actions and to experiment along the way. According to their achievement, students receive experience points in a gamified environment. Professors need to adjust to a new era of teaching and refine their practices and learning philosophy. They need to be able to use virtual platforms with ease, as well as to engage with their students in order to determine and satisfy their needs.

## 1 INTRODUCTION

The COVID-19 disease affected the education systems worldwide. The traditional educational practices have been modified due to the restrictive methods which were imposed. This situation also paved the path towards the introduction of digital learning everywhere around the world. There are also some weaknesses which appear like gaps in information exchange, an environment which does not always suit the most adequate learning conditions, along with equity.

During the last period of time, e-learning digital tools facilitated learning. As the professors and students adapt, the whole study process needs to be monitored and supported, such that the outcomes are positive. There does not exist one pedagogy which would suit online learning. While the students stay at home, they are exposed to psychological and emotional distress which affects their productivity.

The exposure to information and communications technology (ICT) needs to be studied based on several means. Platforms such as Zoom, Microsoft Teams, Google Classroom have been used for educational purposes [1]. Knowledge assessment can be done using quizzes and submitted assignments. There are also classes when students access articles, videos and YouTube links before the lessons [1]. The advantages of online learning are determined by a low Internet bandwidth, some families do not afford themselves to facilitate the access to Internet for their children. A study done in Netherlands showed that almost no progress was done while learning at home for the students with disadvantages situations [2].

The current paper presents how the best learning strategy can be chosen when studying online by analyzing the results coming from an electroencephalogram headset. The brainwaves triggered by the headset electrodes and are interpreted numerically in terms of focus, stress, relaxation, engagement, excitement and interest, along with the mood of the user. The values are entered by the user via a developed web application. In the same application, information about daily activities, learning and evaluation are also provided by the user. Three ways of studying are tested as the user has the headset placed on his/her head, namely text; text and illustrations, including charts; and video. In Section 2 is outlined the proposed system which was designed to make the students more aware of their learning process engagement and inner state during the time of the COVID-19 pandemic, based on the results obtained by using an electroencephalogram (EEG) headset. Section 3 describes the results which were triggered after testing the proposed system by using three kinds of learning materials, namely text; text and illustrations; and video. The last section presents the conclusions.

## 2  METHODOLOGY

Given the shift from traditional to online education, student satisfaction and teaching quality are two directions of interest to evaluate the outcomes. A study done during the COVID-19 pandemic on 162 medical students from Saudi Arabia showed that 82% of them were satisfied to utilize web video conference as a teaching facility [3]. Another study was done on 544 business management students from several Indian universities and it was proved that the study quality of the instructor, course design, fast feedback and student expectation impact their satisfaction [4]. 928 Chinese university students answered to a survey regarding the satisfaction with online learning platforms and the outcomes show that their own computer efficiency, along with the ease of use and usefulness of platforms greatly influenced the results [5]. Similarly, 425 undergraduate students from Hong Kong sustained that their satisfaction greatly depends on assessment methods, structured lessons and interactive learning [6]. All these characteristics need to be taken into consideration when providing a user platform [7], as well as an e-assessment [8].

The process of teaching involves a relationship between work and interpersonal exchange. Mood, focus, stress, relaxation, engagement, excitement and interest are factors which should be considered. They can be determined by using neural oscillations. Not only the environment influences a person's neural activity, but also the interaction with other individuals. Videoconferencing was found to affect mirror neurons, attention, interbrain neural oscillations, all of which affect the cognitive processes [9]. Regarding emotional analysis, a study was done in China to evaluate an online education platform based on the comments provided by users [10]. The comments were analyzed to determine student satisfaction using a structural equation and a back propagation neural network. The results showed that technical problems regarding the platform cannot be neglected and the teaching interaction should be improved by asking questions to the professor when problems appear, as well as student should be allowed to record the learning session. They should also assist to real-time lectures.

The proposed system is designed to analyze how students maintain their engagement. This is studied with the help of neuroscience. Mood, focus, stress, relaxation, engagement, excitement and interest levels are interpreted while trying several teaching approaches, such as inquiry-based learning, game-based learning and personalized learning.

The devices which were used in the proposed system are an Emotiv Insight headset [11], along with a computer, as depicted in Fig. 1. The headset sends the captured waves to the computer via Bluetooth.

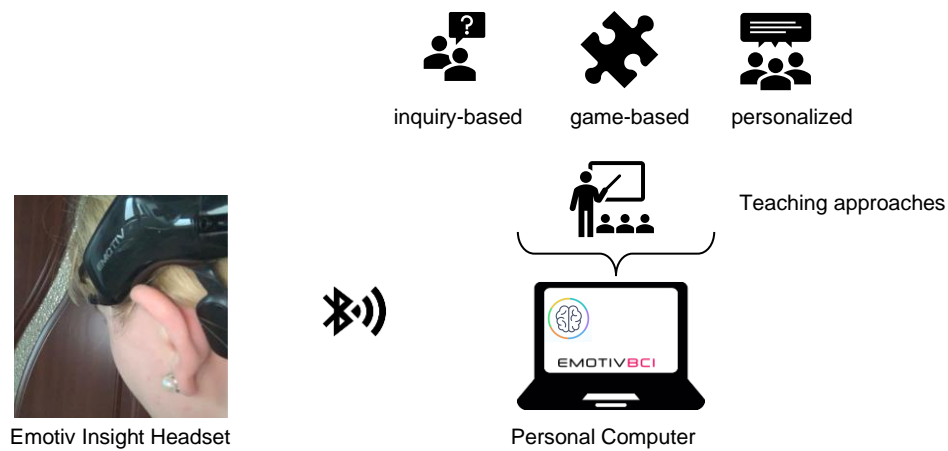

*Figure 1. System architecture*

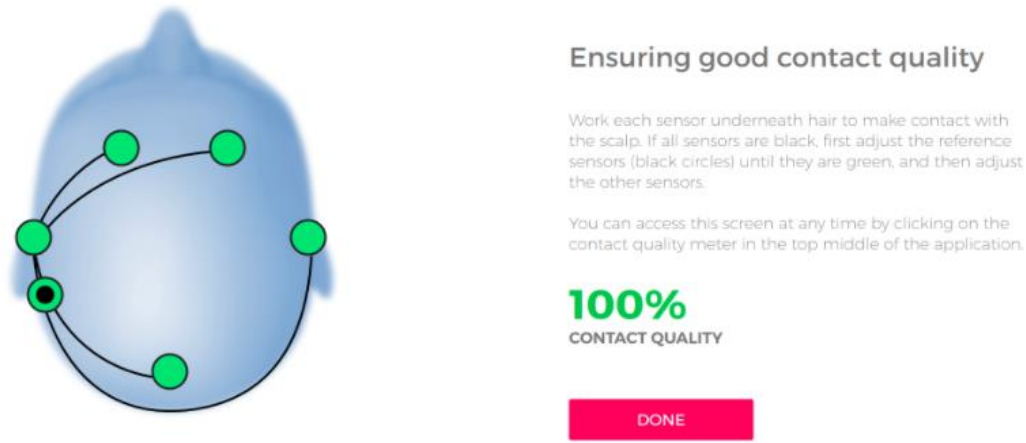

*Figure 2. Sensor contact quality verification*

The computer used in the study has the EmotivBCI [12] program installed and it helps to check if the headset it placed correctly, as well as it indicates the resulted contact quality percentage, as in Fig. 2. The headset contains five electrodes which are depicted using a circle, respectively AF3, AF4, T7, T8 and Pz. If the connectivity is good, the color is green. The circle is a black point in it is in fact the place corresponding to the processor. On the user's forehead are placed the AF3 and AF4 electrodes. AF3 observes the left frontal region activity which controls language related movement [13], while the AF4 electrode monitors the right frontal region activity that corresponds to non-verbal abilities. The signals corresponding to non-verbal emotions greatly influence the likelihood of developing illnesses like depression, anxiety and bipolar disorders, psychopathy [14]. The analysis of sentiments via text was also found to determine changes in heart rate measurements, such that a great attention should be paid to them [15]. The electrodes noted with T7 and T8 are found at the back of the user's ear and head and they are responsible for monitoring activity of the temporal lobe. This lobe handles the processing of auditory information and encoding of memory [16]. The last electrode, Pz, is placed on the top of the user's head and it has the function to have the headset better fitted.

The developed application uses the brainwaves triggered by the Emotiv Insight headset to generate numerical values for several parameters, namely focus, stress, relaxation, engagement, excitement and interest. The values are generated and displayed automatically on the graphical user interface.

The 25 students from the class of Microcontrollers taught in French, at University Politehnica of Bucharest, Romania, used an online simulation environment while studying about Arduino and Raspberry Pi boards. At the same time, the professor did wear the Emotiv Insight headset to determine which teaching method is more adequate. The first teaching approach which was selected was the inquiry-based learning. It has been found that if a lesson embeds inquiry activities, complex problem-solving abilities can be fostered due to the analysis in planning and preparing the steps to solve a problem [17]. This strategy is also part of the PISA methodology when in order to solve a complex problem, it is started by identifying related problems, after which the problem is characterized and comprehended [18]. The next step is to solve the issue and to aid the reflection and communication of the potential solution.

The second studied teaching approach was game-based due to the benefits of improving retention and knowledge transfer in a motivated manner. According to a study, non-native speakers, Germans, proved to have better comprehensive and reading skills in English when educational games were included in their learning process [19]. The results showed that the cognitive load was lower at the beginning of the test, but it increased and stayed at the same level until the end of the trial.

The third teaching approach was personalized learning where the focus is put mostly on student input. The theoretical grounding is not very consistent, such that students can choose the design for a certain situation which is described. The pace of learning is also important in this teaching manner. The students choose and control the steps in developing the solutions for their assignments. The teacher has the role in guiding them during their personal journey. This pragmatic approach leads to an increased engagement from the student side. Personalized learning was found to enhance the development of learner characteristics and to encourage students to take ownership of the learning process [20].

## 3  RESULTS

The study of the system which was outlined in the previous section is based on brain waves which were analysed based on an Emotiv Insight headset, while applying three learning approaches during the class of Microcontrollers taught in French, at University Politehnica of Bucharest, Romania. The headset has five electroencephalogram electrodes that monitor the brain activity via waves. The MyEmotiv APK has been used for the desktop application on Windows [21]. The graphical user interface can be visualized in Fig. 2. The metrics which are displayed have the numerical values between 0 and 100.

The first monitored parameter is stress. A low to medium value of stress signifies an improvement in productivity, while high levels indicate a destructive state which can have consequences for a long period of time over health and wellbeing. High values indicate feelings corresponding to overwhelming and fear of negative outcomes. As illustrated in Fig. 2, the values of stress were under the threshold of 50 throughout the period of teaching, meaning that the situation was kept under control. The stress level increased gradually at the beginning of the monitored period when a practical example was presented, then it was explained, such that the stress decreased, but remained lightly at the same level. The next moment which was captured corresponds to the sequence of questions, when the increase of stress occurred.

The following monitored parameter is engagement which corresponds to alertness and consciousness. Immersion and attention play a key role in this case. Beta waves increase, while alpha waves get attenuated [22]. The value corresponding to engagement is directly proportional to attention, amount of work and concentration. According to the captured data, engagement has the evolution similar to stress, reaching the peak at the end of the practical example and rising again during the questions and answers part of the exposed teaching process.

Another metric is the one for interest, where a low value corresponds to an aversion towards the task, while a high value indicates an affinity. The values belonging to the middle of the range indicate a neural state. As for stress and engagement, in the case of interest, it increases while presenting the practical example, and there is another increase when a question is put. Throughout the whole monitored period which is illustrated in Fig. 2, the values of interest were greater than the average value, 50, denoting that immersion was performed during the teaching activity. Similar to interest is excitement which indicates an activation of the sympathetic nervous system [23]. During the test, excitement reached a peak during the practical example, when the outcomes of the exercise were presented. The excitement slowly decreased afterwards.

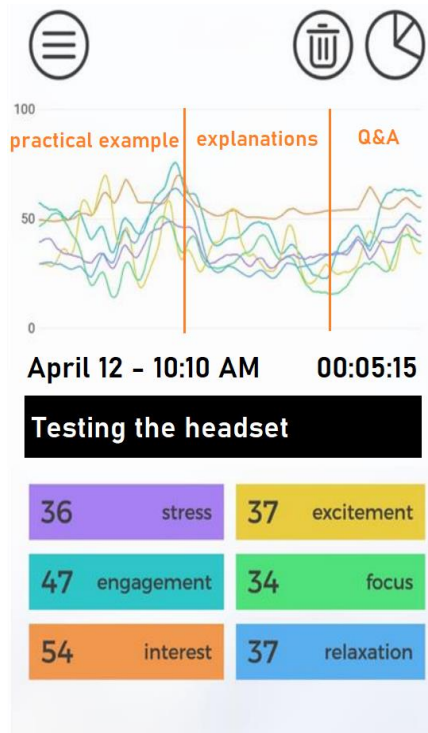

*Figure 2. Channels of the EEG waves during the teaching process*

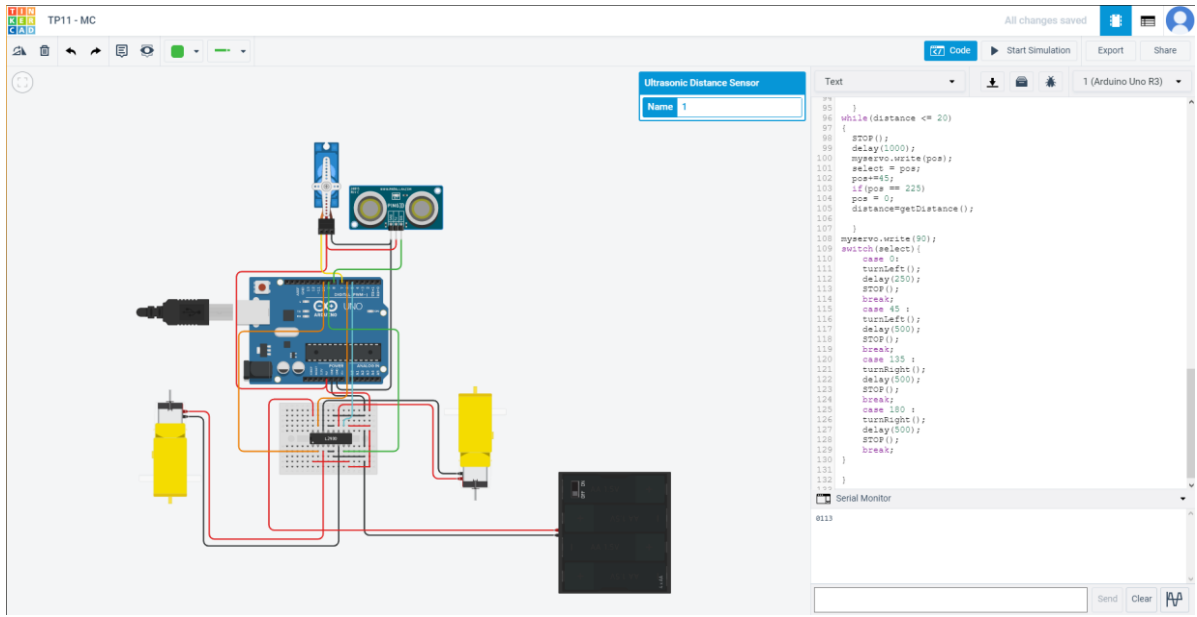

*Figure 3. Gamified environment*

Focus is another metric which can be captured via the Emotiv Insight headset generated brainwaves and it is a measure for determining the attention for a task. Deep attention is indicated by high values, while distraction is denoted by low values. The values for focus which were triggered during the performed study had a similar behaviour as the ones for engagement, showing that focus and engagement are closely connected. The last monitored metric is represented by relaxation that outlines the capability of the user to recover after being exposed to intense concentration. Mindfulness and meditation greatly help in obtaining higher relaxation values, being a low-cost and beneficial solution to support anxiety during the pandemic, mostly now, when education is delivered via online means [24]. During the presented test in Fig. 2, the relaxation values varied in time, the wave being similar to the one for the engagement measurements, but at a difference of about 20 units, proving that the subject is well known and the user feels confident while teaching.

The laboratory assignments were given to be solved using a free web application, Tinkercad [25], where students were able to solve and simulate various sensors, along with code writing, in a ludic manner. Fig. 3 represents the solution to one of the laboratory assignments which the students had to solve based on multiple components, namely positional microservo, Arduino Uno R3, gearmotor, h-bridge motor driver, batteries and an ultrasonic distance sensor.

Three laboratories involved the usage of the Emotiv Insight headset to monitor the variation of the metrics (focus, stress, relaxation, engagement, excitement and interest) and the three types of learnings. For every laboratory, the notions from the course are firstly fixed by using the inquiry-based learning. The next stage was the game-based learning when the students had to use the Tinkercad website to develop the circuit based on the provided schema. For the last exercise of the laboratory, the students had to apply the personalized learning, because they had to come up with their own vision of resolving the problem.

The results triggered after the system was tested during three laboratories of Microcontrollers on 25 students, are illustrated in Fig. 4. The evolution of the monitored parameters, namely focus, stress, relaxation, engagement, excitement and interest, are monitored throughout the three tests, each one of them consisting of inquiry learning, followed by game and personalized learning.

For the first test, during the inquiry period, interest (60) and engagement (55) had the highest values, while the other monitored parameters had the values around 33, meaning that the teacher concentrated during the teaching process, but kept a calm attitude. The next phase, corresponding to game learning, did lead to attaining values over the average value for all six metrics, such that the learning became more demanding, but the situation was kept under control, as the value of relaxation increased compared to the previous phase. The value of stress also increased because the solution is developed in a ludic manner and the resulted circuit has to work. For the last phase of the first test, it is observed that interest reaches a peak value of 75, along with engagement (68), because the accent is put on creativity.

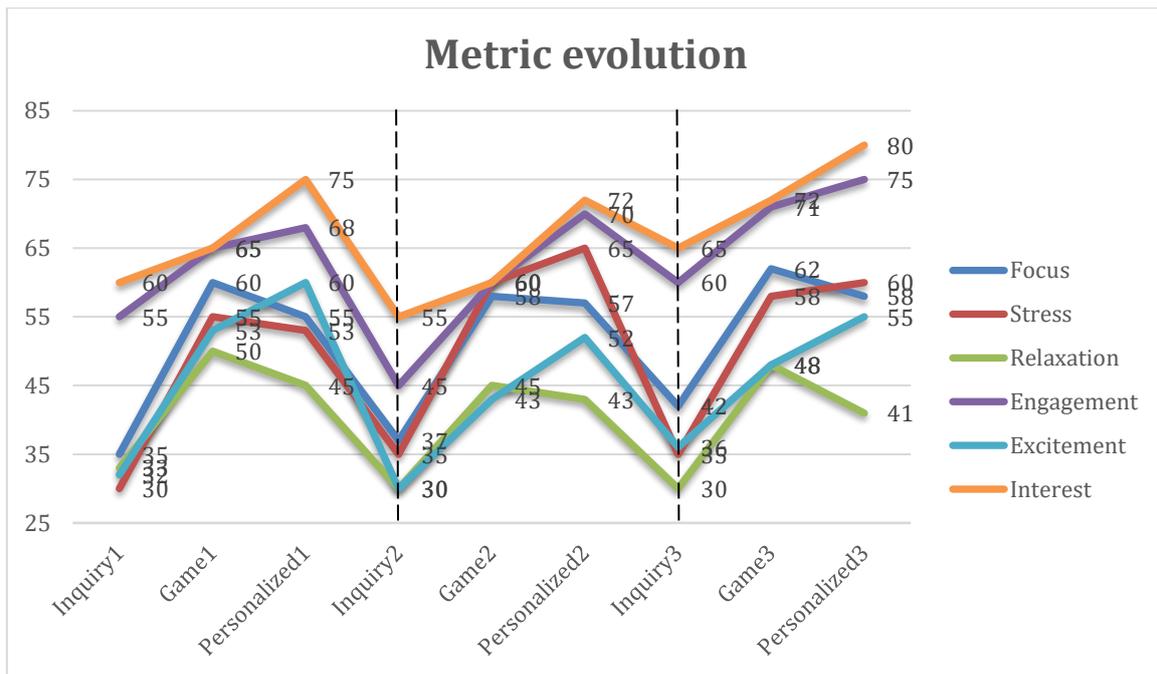

*Figure 4. Studied laboratories outcomes based on the observed metrics*

In contrast, the second laboratory had all the triggered metrics lower than during the first laboratory, the context of it being easier than before, but the stress and engagement values increase because attention should be captured. This also caused a decrease in excitement, but the interest remained high. Relaxation decreased also due to the discomfort of waiting for feedback from the 25 students.

The last studied laboratory was more complex, high interest (65) and engagement (60) being present since the beginning. The students cooperated better during the game learning phase, which did lead to greater values for interest, engagement and focus, than in the previous two tests. The personalized learning phase of the third studied laboratory had as outcome the greatest satisfaction as professor based on the learning experience, as the maximum value for interest (80) and engagement (75) were reached. Focus (58) and excitement (55) were also higher compared to the previous two trials, such that the feedback from students was positive.

All the triggered values by using the Emotiv Insight headset can be utilized to improve the content of the laboratories for the class of Microcontrollers. All the three laboratories were taught in a non-native language, French, such that it is more demanding for the teacher, as well as most of the students who are of Romanian nationality. Teachers, professors and students can utilize the triggered values in order to improve the connection between them, mostly during online learning, as during the three trials which were performed during the spring of 2021 when the pandemic of COVID-19 was still ongoing. Technology and the Tickercad simulation platform greatly helped the process of transferring information towards students from different places around the world, being registered in the third year, at the Faculty of Engineering in Foreign Languages, University Politehnica of Bucharest, Romania.

The best learning method proved to be the game-based learning, which also exists as part of the personalized learning, because students become creative and validate their proposed solutions by playing with the virtual assembled components. All the developed skills during the learning process are essential for next generation software engineers. Their mindset is trained and validated in a virtual environment, where they collaborate with their colleagues, by working in groups. The grades were given based on their experience outcomes in the gamified Tinkercard environment.

## 4 CONCLUSIONS

The proposed system represents a promising solution in the context of the COVID-19 pandemic. The professor is aware of the evolution of focus, stress, relaxation, engagement, excitement and interest during three learning methods, involving inquiries, games and personalized manners. Critical moments can be identified, such as when attention is lost from students, or when they do not answer to certain questions of various difficulties.

The inquiry learning proved to be useful in fixing the notions which the students did not remember or know well. Game learning included better engagement and interest value than compared to inquiry learning, because the process included a ludic manner when problems were solved step by step, constructing the solution out of hardware components and code writing. The personalized learning was the most rewarding learning method because the students used game learning and they have developed their creativity, increased motivation and they focused on reaching the final desired result. The virtual platform performed the overall simulation of the system, even if the students were physically apart from each other.

The current paper outlined how a teaching process can be monitored even when professors are not physically present in a classroom due to the COVID-19 pandemic. The obtained results which were triggered via the Emotiv Insight generated brain waves, showed that game learning and collaborative teaching play an essential role in training the future generations of engineers. The vision of learning gets changed based on the triggered result and the manner of teaching can be revaluated by the professor, as she/he is the one having access to the outcomes. At the end of the test, the professor can adjust her/his approach while teaching and it has been confirmed that virtual simulation platforms can be successfully used, such that students can be motivated and reach their expectations at the end of the course.

## ACKNOWLEDGEMENT

The article was financed by the University Politehnica of Bucharest, Romania, through the project "Inginer în Europa", in online system, registered at ME under no. 140/GP/19.04.2021.